\documentclass[11pt]{article}
\usepackage{moriond,epsfig}
\usepackage{graphicx}

\def\Journal#1#2#3#4{{#1} {\bf #2}, #3 (#4)}

\def\NIMA{{\em Nucl. Instrum. Methods} A}

\def\PLB{{\em Phys. Lett.}  B}
\def\PRL{\em Phys. Rev. Lett.}

\def\JCAP{{\em JCAP}}

\def\be{\begin{equation}}
\def\ee{\end{equation}}
\def\bea{\begin{eqnarray}}
\def\eea{\end{eqnarray}}

\newcommand{\Co} {\mbox{$ ^{57}{\rm{Co}}$}~}

\newcommand{\GeVcc}    {\mbox{$ {\mathrm{GeV}}/c^2                           $}}

\newcommand{\hetrois}    {\mbox{$ ^{3}{\mathrm{He}}                            $}}
\newcommand{\tritium}    {\mbox{$ ^{3}{\mathrm{H}}                            $}}
\newcommand{\xe}    {\mbox{$ ^{129}{\mathrm{Xe}}                            $}}
\newcommand{\ger}    {\mbox{$ ^{73}{\mathrm{Ge}}                            $}}

\begin{document}
\vspace*{4cm}
\title{  MIMAC-He3, \\A PROJECT FOR A MICRO-TPC MATRIX OF CHAMBERS
OF HELIUM 3 FOR AXIAL DIRECT DETECTION OF NON-BARYONIC DARK MATTER}
\author{E. MOULIN and D. SANTOS}

\address{Laboratoire de Physique Subatomique et de Cosmologie,CNRS/IN2P3
and Universit\'e Joseph Fourier\\
53, avenue des martyrs, 38026 Grenoble Cedex, France}

\maketitle\abstracts{MIMAC-He3 (MIcro-tpc MAtrix of Chambers of
Helium 3) is a project for direct detection of non-baryonic dark
matter search using \hetrois \ as sensitive medium. The
priviledged properties of \hetrois \ are highlighted. The double
detection : ionization and track projection, is explained and
rejection evaluated. A phenomenological study in effective SUSY
models has been to investigate the MIMAC-He3 complementarity with
respect to existing Dark Matter detectors. }

\section{Why Helium 3 ?}
As reported elsewhere~\cite{dm2000,nima2000,4micc}, the use of
\hetrois \ is motivated by its very appealing features for
non-baryonic Dark Matter search compared with other target nuclei.
\hetrois \ being a 1/2 spin nucleus, a detector made of such a
material will be sensitive to the axial interaction with WIMPs.
For massive WIMPs, the maximum recoil energy depends very weakly on
the WIMP mass as the \hetrois \ nucleus is much lighter
(m(\hetrois)=\,2.81\,\GeVcc). Therefore, the energy range in which
all the sought events fall, is higher-bounded by $\sim$ 6 keV.
The recoil energy range needs to be studied from energy
threshold up to $\sim$ 6 keV. This narrow range represents a key
point to discriminate these rare events from background. In
addition, the \hetrois \ presents the following advantages with
respect to other materials :
(i) a very low Compton cross section to $\gamma$-rays reducing
by several orders of magnitude the natural radioactive background,
(ii) no intrinsic X rays,
(iii) the neutron signature : the
capture process, $\rm n + \hetrois \rightarrow p + \tritium +
764\, keV$, gives a very clear signal for neutron rejection.
The extremely low sensitivity to $\gamma$-rays and the
possibility to detect events in the keV range ($<$ 6 keV) have
been demonstrated by the electron conversion \Co detection
recently reported~\cite{nima}.
The property (iii) is
a key point for dark matter search as neutrons in underground
laboratories are considered as the ultimate background. Careful
simulations and measurements of the neutron production induced by
high energy muon interaction in the shielding are
compulsory~\cite{kudryavtsev,shielding}. The work developped on
MACHe3 prototype concerning the simulations of the interaction
 of the electromagnetic radiation and cosmic particles and the rejection estimation
 based on the properties above mentioned~\cite{nima2000,4micc} is applied to a new TPC detector offering
the electron-recoil discrimination.

\section{The Micro-tpc}
\begin{figure}[!ht]
\begin{center}
\epsfig{figure=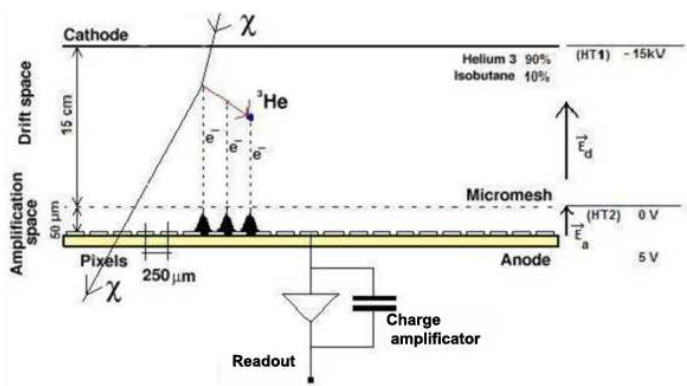,height=5cm}
\caption{\label{fig:ModuleMIMACHe3}A typical chamber of MIMAC-He3
filled with gaseous \hetrois. Ionization electrons induced
produced by recoil event are drifted toward the anode plane. The
pixellised anode allows to recover a 2D projection of the track of
the recoil event.}
\end{center}
\end{figure}

\begin{figure}[!ht]
\begin{center}
\includegraphics[width=10cm,height=5cm]{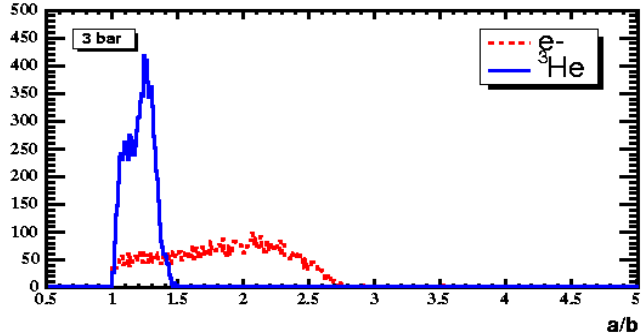}
\caption{Histograms of the a/b distribution for recoils (solid line) and
electrons (dotted line) releasing 4.2 keV in the ionization channel at 3
bar pressure. \label{fig:Ratio}}
\end{center}
\end{figure}

The micro temporal projection chambers present the required
feature to discriminate electron-recoil events by the double
detection of the ionization energy and the track projection in the
anode plane. A schematic view of a typical chamber is sketched on
figure~\ref{fig:ModuleMIMACHe3}. A nuclear recoil induced by a
WIMP scattering off \hetrois, releases ionization electrons along
its track which are drifted towards the amplification space where
they are amplified in an avalanche process. In order to get the
electron-recoil discrimination, the pressure of the TPC should be
such that the tracks of the electrons having energies lower than
4.2 keV can be resolved with respect to those corresponding to the
recoils of the same energy, being the total one convoluted by the quenching factor.
\begin{figure}[!ht]
\begin{center}
\epsfig{figure=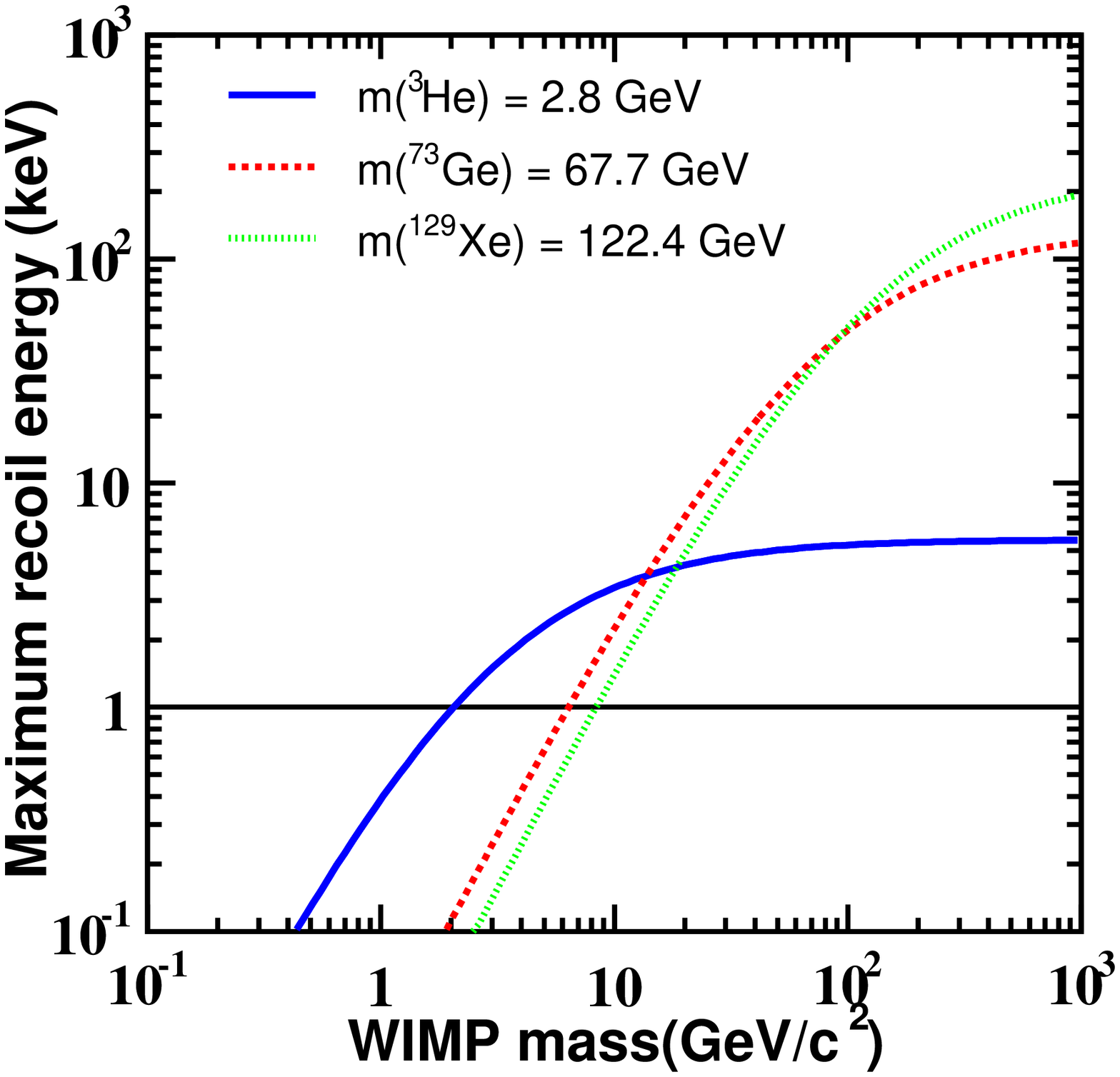,height=6cm}
\hspace{0.5cm}
\epsfig{figure=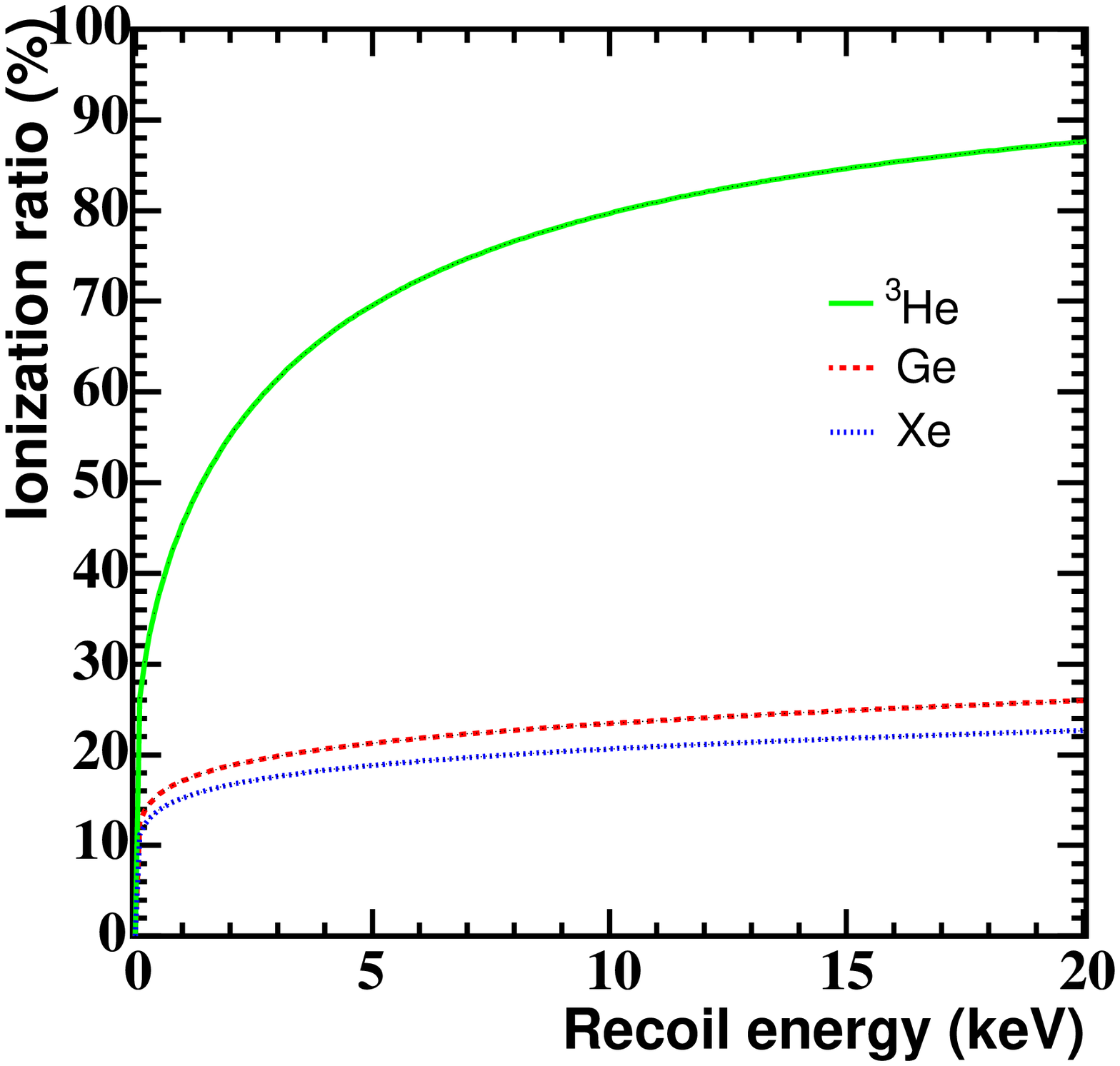,height=6cm}
\caption{The plot on the left hand
side shows the maximum recoil energy of the target nucleus versus
the incident WIMP mass for different target nuclei. For \hetrois,
the energy range where all the sought events have to fall in is
higher-bounded by 5.6 keV. In the case of heavier nuclei such as
\ger \ and \xe, the energy range is much wider. The right had side
plot exhibits the ionization ratio predicted by Lindhard
calculations for different sensitive media. For \hetrois, up to
70\% is expected to be released in the ionization channel.
\label{fig:IonizationRatio}}
\end{center}
\end{figure}
No measurement on the quenching factor (QF) in \hetrois \ have
been reported. However, the quenching factor based on Lindhard
calculations~\cite{lindhard} has been estimated in order to
quantify the amount of the total recoil energy that is released in
the ionization channel. The quenching factors for different pairs
of nuclei (\xe, \ger \ and \hetrois) of the same atomic and mass
numbers are plotted on figure~\ref{fig:IonizationRatio}. The Ge
curve is in good agreement with measurements as shown
in~\cite{shutt}. From the \hetrois \ curve, up to 70\% of the
recoil energy in the energy range of interest is going through
ionization. For 6 keV total energy recoil, 4.2 keV is released in
the ionization channel. In order to measure the QF in \hetrois \
in the keV range, an ion source has been designed at LPSC to
accelerate helium ions to be coupled to the micro-tpc chamber.
Accelerated ions by the source will pass through a thin
polypropylene foil that will neutralize them before entering in the
prototype chamber. The measurement of the energy of the \hetrois \
atom entering in the chamber will be made by a time of flight
measurement. For more details, see~\cite{mimache3}.
In order to characterize the distribution of pixels on the anode
plane produced by different kind of trajectories, we have defined
the ratio between perpendicular symmetry axis of the pixel
distribution (a/b) where a is the larger axis of the distribution.
Simulations have been performed at different energies of electrons
with Geant4~\cite{g4} and recoils with SRIM~\cite{srim} for
different pressures. An isotropic spherical emission of electrons
and recoils at the center of the chamber has been injected as 
input of the simulation. The drift of the ioniziation electrons
has been simulated by Garfield~\cite{garfield} following the well
known distribution characterized by a radius of $\rm D\sim 200 \mu
m \sqrt{L(cm)}$ where L is the total drift in the chamber.
Figure~\ref{fig:Ratio} presents the ratio a/b for electrons and
\hetrois \ releasing 4.2 keV in the ionization channel at 3 bar
pressure.  For recoils, a concentrated distribution around 1
is expected, and for electrons, a much wider one. The rejection of
events using the a/b ratio is a strong fonction of the energy and
the pressure of the chamber, but even at 1 keV, only a small
number of the total events can be confused~\cite{mimache3}.

\section{Supersymmetric Dark Matter search with MIMAC-He3}
\begin{figure}[!ht]
\begin{center}
\mbox{\hspace{-.5cm}\includegraphics[scale=0.5]{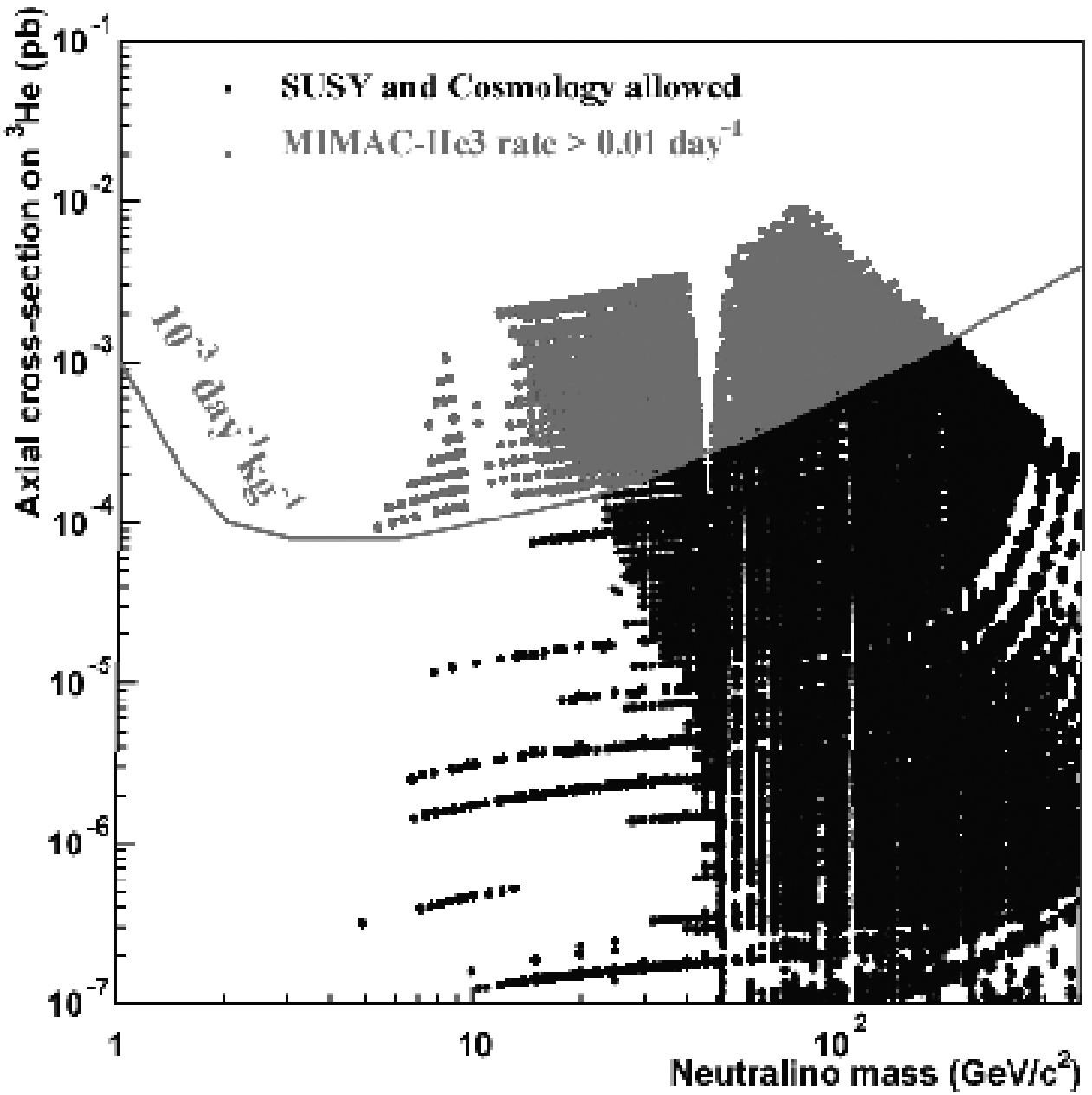}\hspace{1cm}
\includegraphics[scale=0.5]{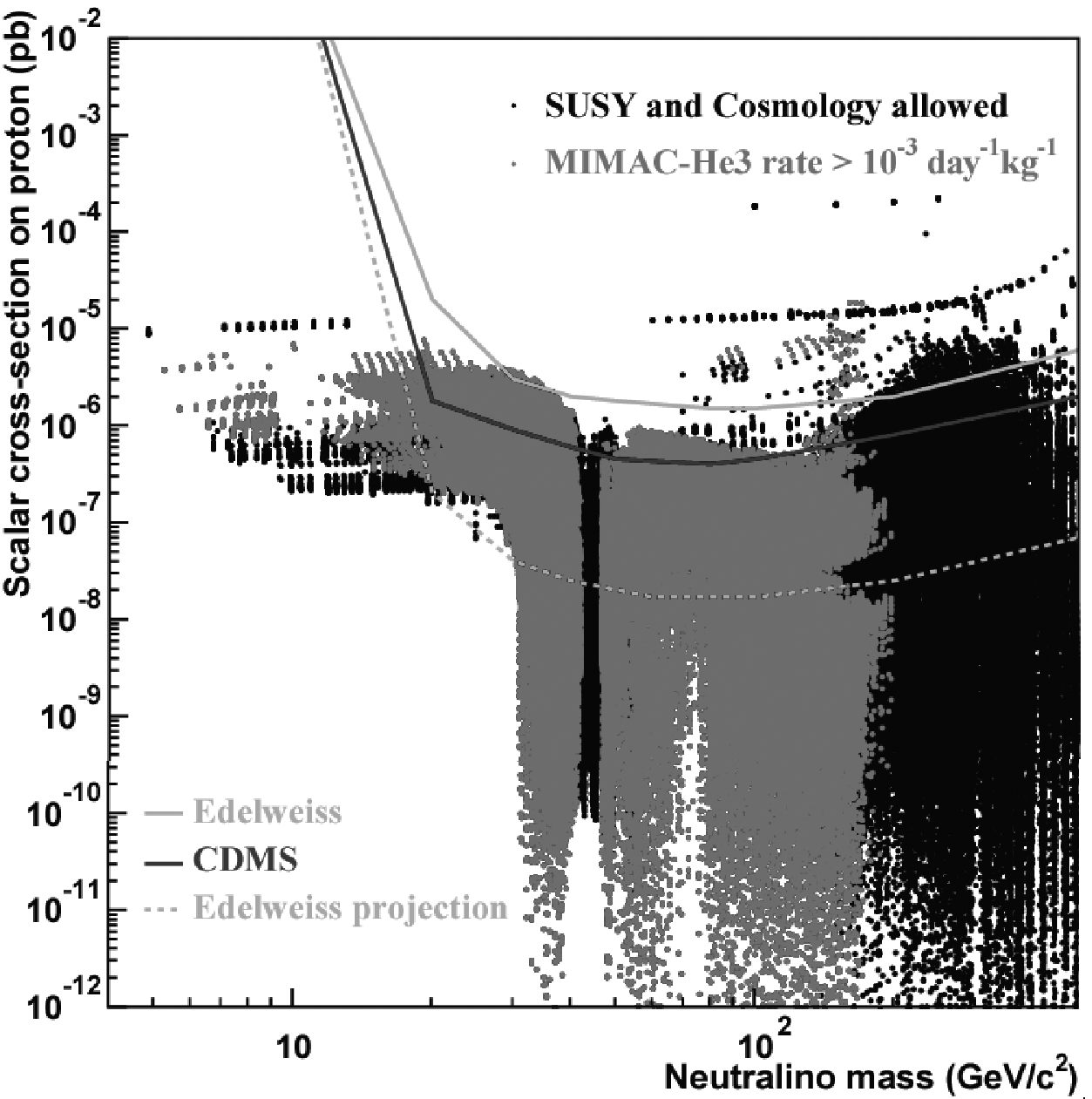}}
\caption{The left side plot shows the axial cross section on
\hetrois \ (pb) \ versus the neutralino mass (\GeVcc). Black points
corresponds to SUSY models satisfying collider as well as
cosmological constraints. Projected exclusion curves for MIMAC-He3
(red solid line) with 10$^{-3}$ day$^{-1}$kg$^{-1}$ background
level is drawn. Models accessible by MIMAC-He3 correspond to red
points. The right hand side plot presents the scalar cross section
on proton (pb) versus the neutralino mass (\GeVcc). Exclusion
curves from CDMS$^{15}$ and Edelweiss$^{16}$
experiments are plotted as well as the projected limit for Edelweiss (dotted line).\label{fig:susy}} 
\end{center}
\end{figure}

A phenomenological study using the DarkSUSY code~\cite{ds} has
been done to investigate the potentiality of MIMAC-He3 for
supersymmetric dark matter search. Within the framework of
effective MSSM models without no gaugino unification at GUT scale,
the sensitivity of the MIMAC-He3 detector to neutralino masses $\rm
M_{\tilde{\chi}} > 6\, GeV/c^2$ via spin-dependent interaction
with \hetrois \ has been evaluated~\cite{susy}. In a large SUSY parameter space, the
axial cross-section on \hetrois \ has been calculated as well as the
$\tilde{\chi}$ event rate in MIMAC-He3 and its complementarity to
direct and indirect detection has been shown~\cite{susy}.
Figure~\ref{fig:susy} presents on the left hand side plot the SUSY
models accessible to MIMAC-He3 (grey points) in the axial cross section on \hetrois
versus neutralino mass plane.
On the right, the same models
in scalar cross section on proton plane are presented. A large number of models accessible to
MIMAC-He3 escapes to ongoing saclar detector sensitivitis and even to their projected
exclusion curves. In the framework of non-universal gaugino
masses, neutralinos can be lighter than $\sim$ 50 \GeVcc. \hetrois \ being a light target mucleus,
MIMAC-He3 will be sensitive to light masses down to $\sim$6
\GeVcc~\cite{susy}.


\end{document}